# Structural Restrictions and Inorganic Nanotubular Growth

Albert Prodan, Herman J. P. van Midden, and Erik Zupanič
Jožef Stefan Institute, Jamova 39, SI-1000 Ljubljana, Slovenia

corresponding author: albert.prodan@ijs.si

Abstract

Inorganic nanotubular growth is briefly discussed. It is argued that such tubes with thicker walls can be grown from strongly bonded multi-layered structures only in accord with some basic crystallographic principles. Examples are given to show that local symmetry and arrangement of nearest neighbors are easily adjusted to new forms, while the corresponding interatomic distances vary within narrow limits only. Consequently, the diameters of such tubes must remain large in comparison with the thicknesses of their walls. For smaller diameters new bulk structures must be formed.

# 1. Introduction

Soon after the discovery of carbon fullerenes (Kroto et al. 1985) and nanotubes (Iijima 1991) transition-metal dichalcogenide ($MX_2$) (Tenne et al. 1992) and dioxide ($MO_2$) (Hoyer 1996) tubular forms were also reported. However, possible forms of inorganic tubes made of strongly bonded multiple layers (Tenne 2006) are in comparison to nanotubes, formed of carbon and the structurally related compounds, subjected to more stringent structural restrictions. While tubular growth is supported by a removal of the rim atoms, the deformation introduced by bending the multiple layers necessarily results in structural constrains, whose importance grows with reduced diameters of the tubes. These restrictions should not be underestimated and solid proofs regarding structure and composition are indispensable to avoid misinterpretations.

Inorganic tubular and fullerene-like structures are nowadays the subject of numerous studies. A few review articles and books (Ram and Govindaraj 2005; Tenne 2006; Yan et al. 2008) were i.a. devoted to their structure and properties. The reported radii in case of $MX_2$ compounds vary largely, from fullerene-like particles with diameters of the order of a few ten micrometers to ordered tubes with nanometer diameters. To the contrary, the tubes made of/in oxides are usually of larger diameters and if prepared by electrochemical methods, e.g. by making holes into continuous layers or into nanorods (Shin and Lee 2008; Yan et al. 2008) their formation will not be mainly dependent on restrictions connected with deformation of the original structure.

A problem in this context may represent the experimental methods used. The structure and composition of individual tubes is usually studied by means of various microscopic techniques, while assembles of tubes are often investigated by x-ray powder methods. Both can be unreliable. Overlapped characteristic lines of the constituent elements may cause problems in case the composition of such tubes is determined by means of x-ray energy dispersive analysis, while oxide nanoparticles and mixtures of phases in particular, like e.g. those of different $M_nO_{2n-1}$ Magnelli phases, will often reveal x-ray powder diffraction spectra composed of rather broad peaks, which are easily misinterpreted.

# 2. Structural Consideration

Growth of any new form, and $MX_2$ nanotubes made of rolled up multiple layers make no exception, must take place in accord with certain crystallographic restrictions. Some structural parameters are easily adjusted to new forms, while others can undergo minor changes only. A quick check will show e.g. that Nb-Nb, Nb-Se and Se-Se distances in various $NbSe_2$ polytypes are of comparable lengths, regardless of the coordination involved (Hulliger 1976). The same is true if structurally and compositionally more diverse arrangements of the same constituent elements are compared, like e.g. $Nb_3Se_4$, $NbSe_2$, $NbSe_3$, and $(NbSe_4)_3I$. This includes comparison of the few considerably shorter double bonds in these compounds. Contrary to the inter-atomic distances the distribution of the nearest neighbors, which form coordination polyhedra in the mentioned Nb-Se compounds, can differ considerably from case to case. The simplest examples are again the polytypes; Nb atoms in all $NbSe_2$ polytypes are coordinated by six Se atoms, distributed either in an octahedral or trigonal prismatic arrangement. The same is true for the mentioned compounds with more diverse compositions. The Nb coordination number in $NbSe_3$ is eight, forming bi-capped trigonal prisms with equilateral bases, while Nb atoms in $(NbSe_4)_3I$ form with adjacent Se atoms eight-coordinated rectangular antiprisms and in $Nb_3Se_4$ six-coordinated deformed octahedra (Selte and Kjekshus 1964; Hulliger 1976; Merschaut 1977; Hodeau et al. 1978).

The given examples show clearly the restrictions, which are to be respected when growth of different forms is considered. A strongly bonded multi-layered $MX_2$ compound can only be transformed into a tube with the same, only slightly deformed) original structure as long as its diameter requires only minor adjustments of the inter-atomic distances. The gain in energy by removing the rim atoms at the edges of the two-dimensional layered structure must more than compensate the increase in elastic energy caused by deformation. Consequently, the diameters of such tubes must remain relatively large in comparison with the thicknesses of the original layers, which form their walls. $MX_2$ and $MO_2$ tubes are in this regard very different from carbon nanotubes and fullerene-like structures, particularly if these appear as single-walled structures. With the discussed restrictions in mind, their tubular growth can beyond any doubt be under certain conditions energetically favorable, but hardly with diameters of the order of nanometers. This requirement is also supported by ab-initio calculations, performed on a number of comparable one-dimensional structures (Verstraete 2003). According to these calculations very narrow tubes, formed of deformed $MX_2$ layered compounds, are between the energetically least favorable possibilities.

## 3. Discussion

Once multi-walled $MX_2$ tubes are considered, different scenarios are possible. What may happen in case a layered $MX_2$ compound is bent into a very narrow tube is shown in Fig.1a,b.

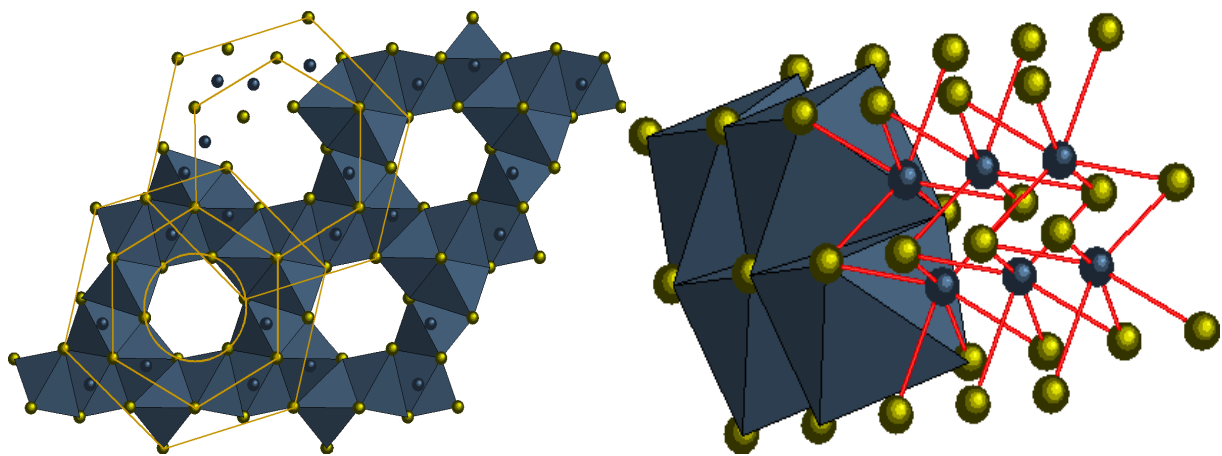

Figure 1. *Possible formation of a tube from a layered $MX_2$ structure: From two types of tubes, shown by larger and smaller hexagons, only the wider ones can appear as isolated tubes, while the narrow ones require too large X-X bond lengths (a). If isolated, the tubes must keep all interatomic distances within acceptable limits, as shown for a pair of octahedral columns, forming the suggested tubes (b). The circle indicates the fixed inner diameter of a tube, yellow balls represent the X and blue balls and octahedra the M atoms, while red rods show the corresponding M-X bonds.*

In agreement with the given arguments, the six-coordinated transition-metal atoms can form with the surrounding chalcogen atoms either octahedra or trigonal prisms, both with similar M-X bond lengths. The inner diameters of such tubes are determined by the number of chalcogen atoms forming their circumferences. These can be neither enlarged nor reduced without changing the number of atoms that form them. Further, since the inter-atomic distances are to be kept within certain limits, isolated tubes with $MX_2$ composition and with radii of the order of a nanometer can only be constructed by filling the remaining empty space with a second ring of M atoms and a corresponding number of outer X atoms. Such a tube with thicker walls is shown in Fig.1a by the larger hexagons, each corresponding to a composition $X_6$(inner)$M_{12}X_{18}$(outer). These tubes are to be compared with the narrow $X_6$(inner)$M_6X_6$(outer) tubes, indicated by the smaller hexagons. Although both, if isolated, correspond to the $MX_2$ composition, the narrow ones cannot stand on their own as individual tubes, because their outer X-X interatomic distances appear unrealistically large. The tubes with thicker walls might in principle be found as individual tubes (Fig.1b), but their inner and outer diameters cannot be varied at this scale continuously. Both are simply determined by the structure. If in spite of that such tubes are experimentally found to be ordered with shorter periodicities, they have to be intergrown. A most probable solution in the discussed case is thus formation of the $M_3X_4$ quasi one-dimensional structure (Selte and Kjekshus 1964), also shown in Fig.1a. It should be noted that contrary to a series of $M_3X_4$ isostructural compounds, which all crystallize in the space group $P6_3/m$ with slightly different parameters (e.g. $Nb_3S_4$: a = 0.95806 nm, c = 0.3375 nm; $Nb_3Se_4$: a = 1.0012 nm, c = 0.34707 nm; $Nb_3Te_4$: a = 1.0671 nm, c = 0.3647 nm), $Mo_3S_4$ was so far reported only as a Chevrel phase (Chevrel 1974). However, its rhombohedral structure, with six Mo atoms forming an octahedral cluster inside a cube with eight S atoms at the corners, can be easily related to the $Nb_3Se_4$ structural type. Thus, a second form may e.g. be stabilized by the presence of a catalyst, present during the growth process, which will be intercalated into the structural hexagonal tunnels. Other solutions are possible, but they will all have to obey the same principles and will likewise result in bulk structures, different from the original layered $MX_2$ one.

Similarly, there are numerous ways to form tunnels with nanometer diameters in oxides as well, which will not require bending of the original structures. The $MO_2$ rutile structure can e.g. be transformed by a periodic rotation of groups of four $MO_6$ octahedra into a hollandite structure with large square tunnels. Again other transformations may involve formation of hexagonal tungsten bronzes with hexagonal tunnels.

## 4. Conclusions

Formation and ordering of inorganic $MX_2$ nanotubes with thicker, strongly bonded walls can take place without changing the original structure only as long as the diameters of the tubes remain large in comparison with the thicknesses of their walls. Below certain critical diameters new bulk structures will have to be formed, with fully determined inner and outer diameters of the intergrown tubes.

**Acknowledgement**

Financial support of the Slovenian Research Agency is kindly acknowledged.